\documentclass[twocolumn,prc,aps,floatfix]{revtex4}
\usepackage[dvips]{graphicx}
\begin{document}
\title{Fast, Runaway Evaporative Cooling to Bose-Einstein Condensation in Optical Traps}
\author{Chen-Lung Hung, Xibo Zhang, Nathan Gemelke, Cheng Chin}
\affiliation{James Frank Institute and Physics Department, The
University of Chicago, IL 60637}
\date{\today}

\begin{abstract}
We demonstrate a simple scheme to achieve fast, runaway evaporative
cooling of optically trapped atoms by tilting the optical potential
with a magnetic field gradient. Runaway evaporation is possible in
this trap geometry due to the weak dependence of vibration
frequencies on trap depth, which preserves atomic density during the
evaporation process. Using this scheme, we show that Bose-Einstein
condensation with $\sim 10^5$ cesium atoms can be realized in $2\sim
4$~s of forced evaporation. The evaporation speed and energetics are
consistent with the three-dimensional evaporation picture, despite
the fact that atoms can only leave the trap in the direction of
tilt.

\end{abstract}

\maketitle

The possibility to manipulate Bose-Einstein condensates (BECs) and
degenerate Fermi gases of cold atoms in optical traps opens up a
wide variety of exciting research; prominent examples include spinor
condensates \cite{stenger1998}, Feshbach resonance in cold
collisions \cite{inouye1998}, and BECs of molecules
\cite{jochim2003, greiner2003}. In many early experiments,
condensates were first created in a magnetic trap and subsequently
transferred to an optical dipole trap. These experiments could be
greatly simplified after direct evaporation to BEC in optical traps
was demonstrated \cite{barrett2001}.  In this paper, we describe a
further improvement on dipole-trap based evaporation, which allows
for runaway cooling without significant increase in trap complexity.


Evaporative cooling proceeds by lowering the depth of a confining
potential, which allows atoms with high kinetic energy to escape and
the remaining particles to acquire a lower temperature and higher
phase space density through rethermalization. Starting from a sample
of precooled atoms in a dipole trap, one can in principle perform
forced evaporative cooling on optically trapped atoms by constantly
reducing the trap depth until quantum degeneracy is reached. This
method has been successful in creating rubidium BEC in a dipole
trap, and has become a critical component in recent experiments on
quantum gases of Cs \cite{weber2002}, Li \cite{ohara2002}, K
\cite{regal2003} and Yb \cite{takasu2003}. In all these experiments,
forced evaporative cooling in the dipole trap is realized by
reducing the intensity of the trapping beam, and consequently also
the restoring forces. In later discussion, we will refer to this
approach as trap-weakening scheme.

Evaporative cooling in optical traps remains one of the most
time-consuming and technically challenging steps in condensate
production. Fundamentally, this is due to the fact that cooling by
weakening the trapping potential inevitably reduces the collision
rate. Here runaway (accelerating) evaporation is essentially
impossible even with perfect evaporation efficiency and purely
elastic collisions \cite{unitarygas}. Within experimentally
accessible times, the trap-weakening method puts a severe limit on
the maximum gain in phase space density one can reach. Several
auxiliary schemes have been successfully implemented in order to
increase the evaporation speed, including the dimple trap
\cite{weber2002} and a zoom lens system \cite{Kinoshita2005}. These
methods often increase the complexity of the apparatus or require
delicate optical alignment or manipulation.


\begin{figure}
\includegraphics[width=1.07in]{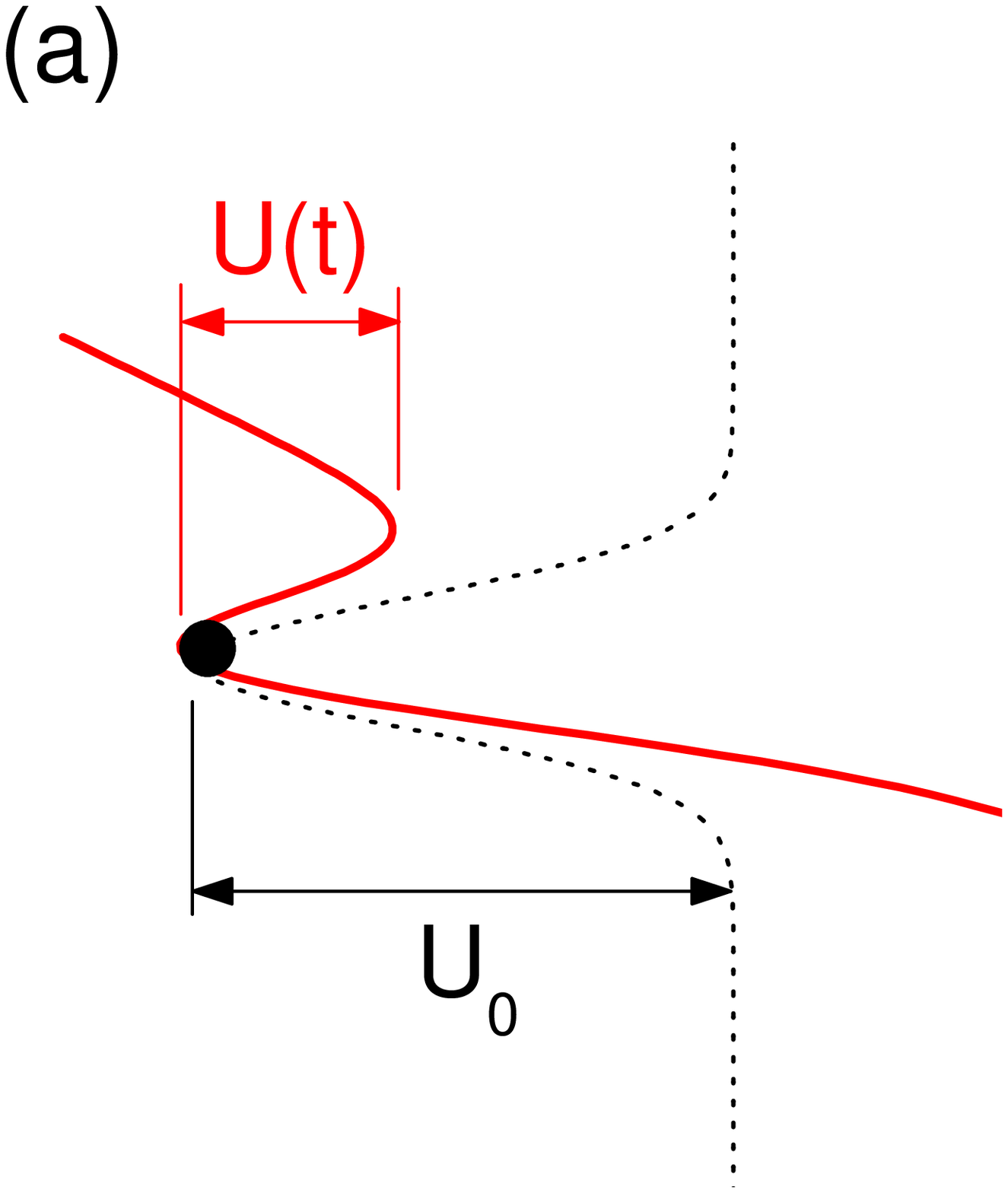}
\includegraphics[width=2.25in]{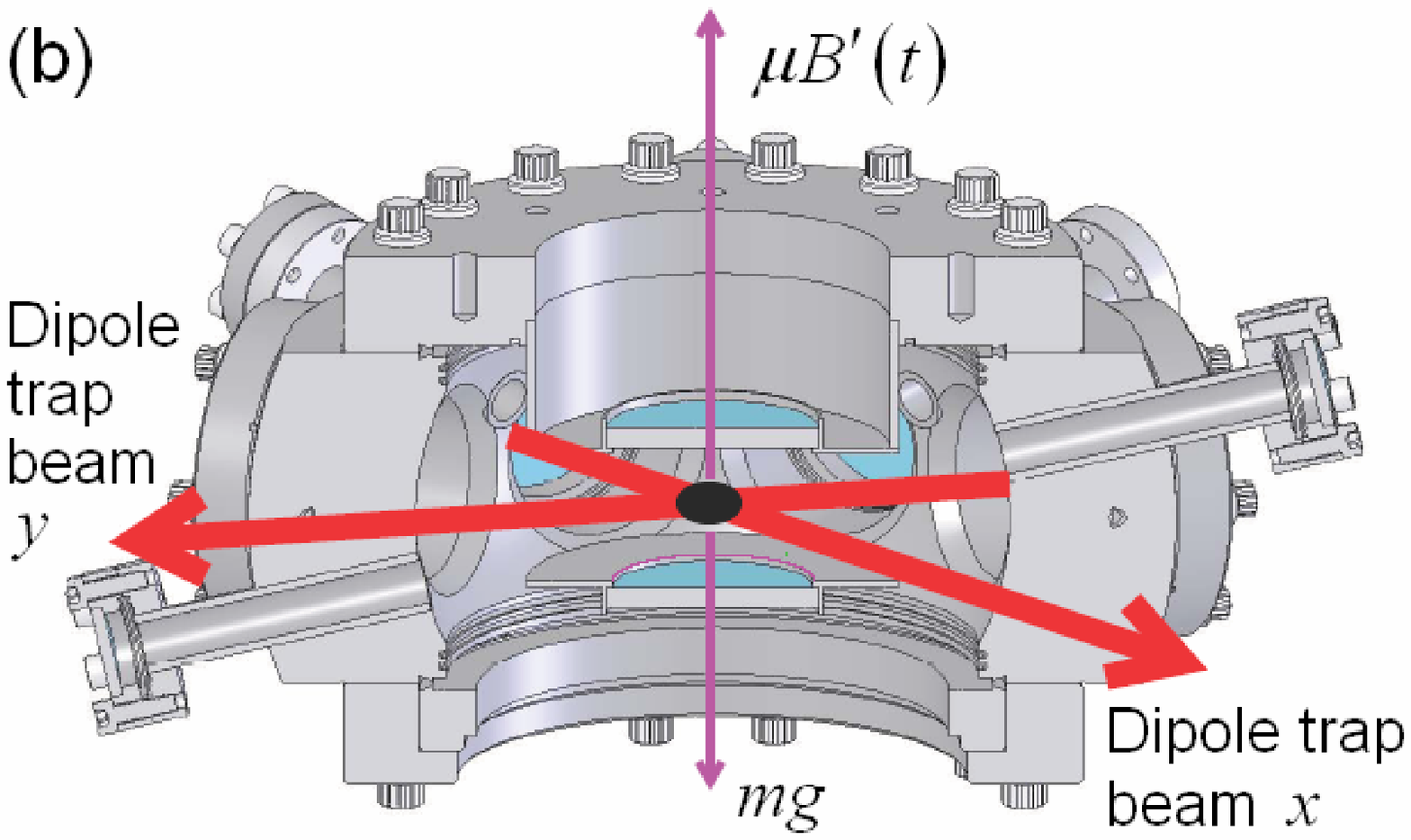}
\caption{(color online) Trap-tilt based evaporation and experimental
apparatus. (a) Trap depth $U$ decreases when an external potential
gradient is applied to the optically trapped atoms. (b) Apparatus
for evaporation of cesium atoms (black dot) in a crossed-beam dipole
trap. A strong, slowly-varying magnetic field gradient $B'(t)$
over-levitates the atoms with magnetic moment $\mu$ against
gravitational pull $mg$ and evaporates them upward. } \label{psd}
\end{figure}

In this paper, we report a new and simple evaporative cooling scheme
which can be immediately implemented in many existing experiments.
Instead of reducing the intensity of the trapping beam, we reduce
the trap depth by applying an external force on the optically
trapped atoms. This trap-tilting method entails only a weak
reduction in confinement strength over a large range of potential
depth and can significantly speed up the cooling process. Using this
method, we demonstrate runaway evaporative cooling in a large volume
dipole trap and reach Bose-Einstein condensation of cesium
significantly faster than previous results \cite{kraemer2004}.
Finally, we comment on the conditions for runaway evaporation in a
tilted trap and investigate the dimensionality of atomic energy
selection in the evaporation.

\begin{figure}
\includegraphics[width=2.25 in]{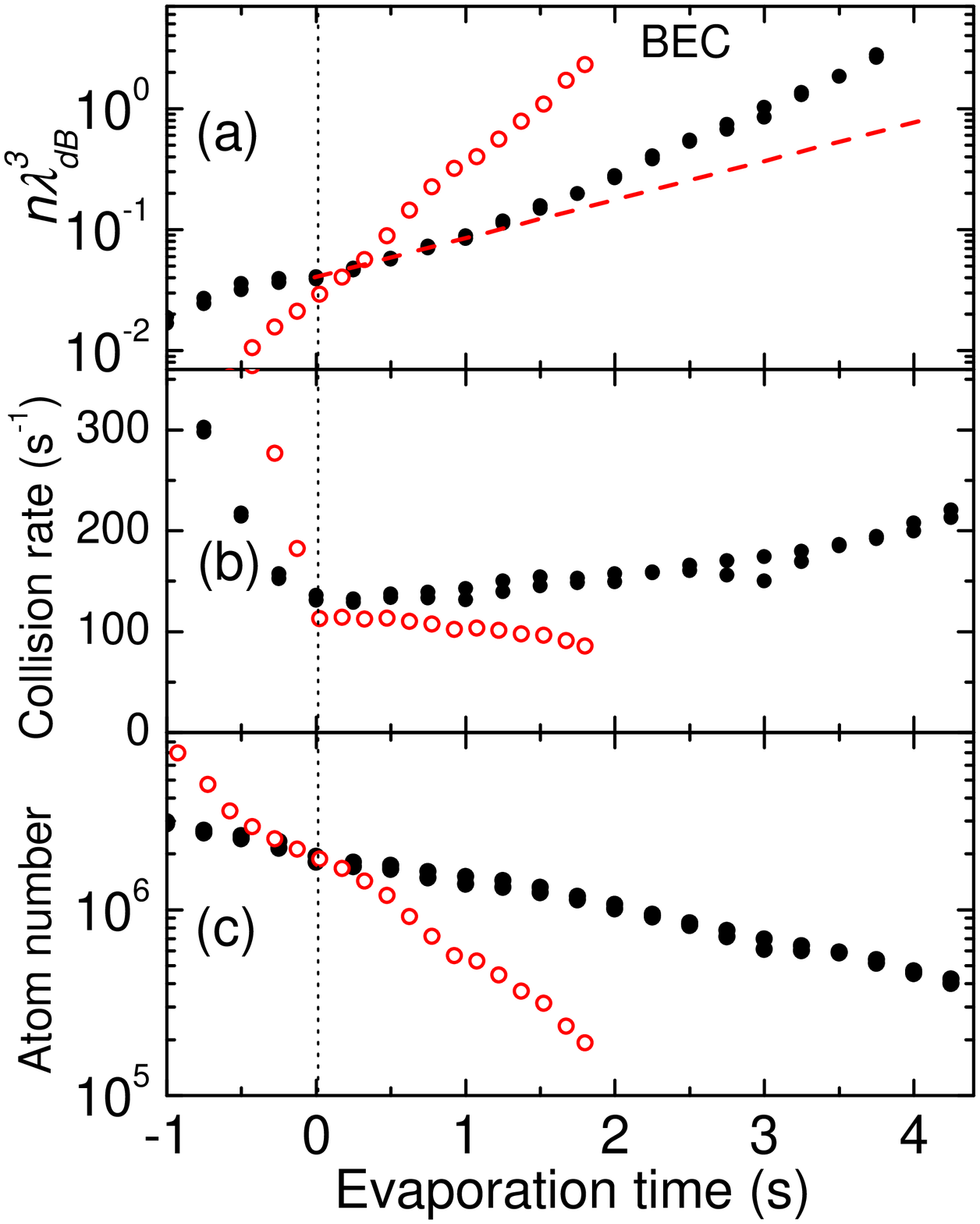}\\
\includegraphics[width=3.1 in]{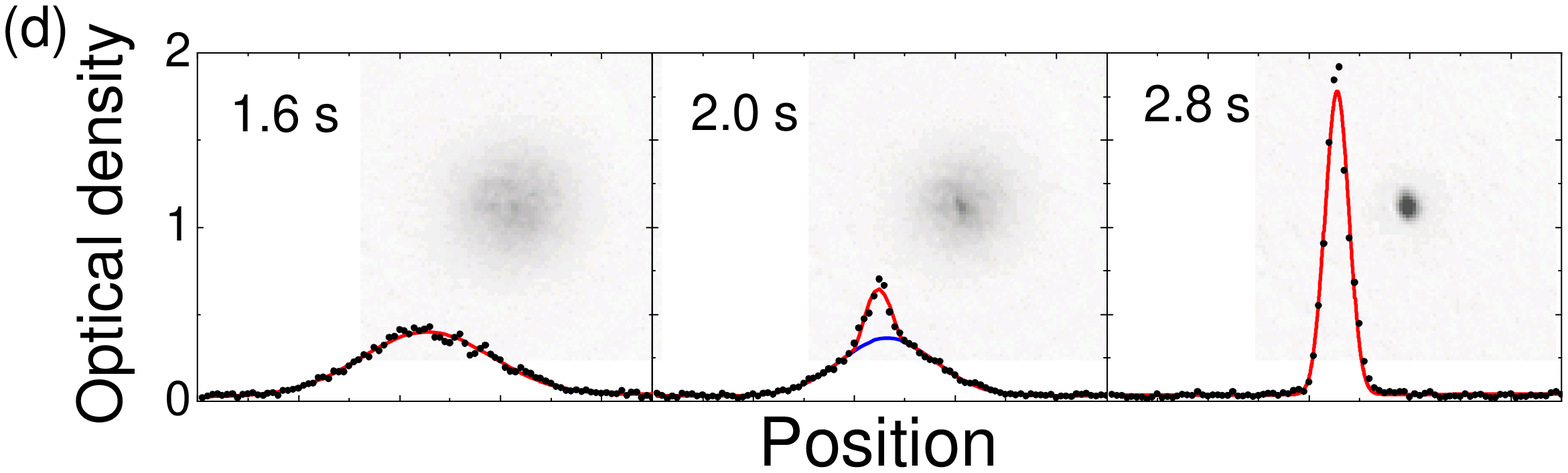}
\caption{Performance of trap-tilting based forced evaporation: (a)
phase space density, (b) collision rate, (c) particle number and (d)
density profile (d). Two evaporation paths: 4~s (solid dots) and
1.8~s (open circles) are shown. The dashed line in (a) shows simple
exponential increase. In (d), time-of-flight absorbtion images and
single-line optical density profiles are taken from the 1.8~s
evaporation path. The expansion time is 70~ms, and the field of view
is 1.2~mm $\times$ 1.2~mm.} \label{nsigmav}
\end{figure}

For this study, cesium atoms are first slowed by a Zeeman slower,
collected in a magneto-optical trap (MOT) for 2~s, molasses
precooled, and finally cooled and spin polarized by degenerate
Raman-sideband cooling (dRSC) \cite{kerman2000} to the lowest
hyperfine ground state $|F=3,m_F=3\rangle$, where $F$ is the total
angular momentum and $m_F$ is the magnetic quantum number; the
apparatus for dRSC follows that in \cite{kerman2000}. A crossed
dipole trap and magnetic field gradient are employed to levitate and
collect the cooled atoms. The dipole trap is formed by intersecting
two laser beams on the horizontal ($x-y$) plane; both beams are
extracted from a single-mode, single frequency Yb fiber laser
operating at the wavelength of 1064~nm, frequency offset by 80~MHz,
focused to a $1/e^2$ beam diameter of 540~$\mu$m (620~$\mu$m) and
intensity of $1.9$~W ($1.6$~W) in the $y-$($x-$) direction. In the
absence of trap tilt, the trapping frequencies near the bottom of
the potential well are
$(\omega^0_x,\omega^0_y,\omega^0_z)=2\pi\times (17,34,38)$~Hz.
During the dipole trap loading process, we switch on a uniform
magnetic field of 58~G in the (vertical) $z$-direction to improve
the atom number following the loading process \cite{weber2002} and
apply a levitating magnetic field gradient of
$B'_c=mg/\mu$=31.3~G/cm, where $mg$ is the gravitational force,
$\mu=0.75\,\mu_B$ is the magnetic moment of the atoms in
$|3,3\rangle$, and $\mu_B$ is Bohr magneton. After 1~s of
thermalization and self-evaporation in the dipole trap, we ramp the
magnetic field to 20.8~G, where three-body loss is minimized
\cite{kraemer2006}, and the field gradient to 37.8~G/cm in 1.85~s
and begin our study on forced evaporation. At this point, which we
define as time $t=0$, there are $N_0=1.9\times 10^6$ atoms in the
trap with a temperature of $T_0=470$~nK, peak atomic density of
$n_0=3.8\times 10^{12}$~cm$^{-3}$, peak collision rate of
$\Gamma_0=$133~/s. The background collision rate are below 1/60~s.

We perform forced evaporative cooling by linearly increasing the
magnetic field gradient $B'$ from 37.8 to 41.4~G/cm in 2.2~s and
then to 43.5~G/cm in another 3~s, which reduces the trap depth from
3.0~$\mu$K to 1.0~$\mu$K and then to 170~nK, as calculated from our
potential model. The magnetic field and dipole trap intensity are
kept constant throughout the process. To evaluate the cooling
performance, we interrupt the evaporation at various times to
measure the particle number $N$, temperature $T$ and trap
frequencies $\omega_{x,y,z}$. Particle number and temperature are
extracted from absorption images taken at low magnetic fields,
following a 70~ms time-of-flight expansion at $B=17$~G to minimize
the collisions and $B'(z)=B'_c$ to levitate the atoms. Trap
frequencies are measured from small amplitude oscillations of the
atomic momentum by abruptly displacing the trap center. Peak phase
space density is calculated from $\phi=n\lambda_{dB}^3$, where
$n=N\omega_x\omega_y\omega_z(m\lambda_{dB}/h)^3$ is the peak atomic
density, $\lambda_{dB}=h(2\pi m k_BT)^{-1/2}$ is the thermal de
Broglie wavelength, $k_B$ is the Boltzmann constant and $h$ is the
Planck constant. Collision rates are calculated as
$\Gamma=n\langle\sigma v\rangle$, where the elastic collision cross
section is $\sigma=8\pi a^2$, scattering length at 20.8~G is
$a=$~200~$a_0$ \cite{chin2004}, $\langle v\rangle=(16k_BT/\pi
m)^{1/2}$ is the mean relative velocity.

After 4~s forced evaporative cooling, we observe Bose-Einstein
condensation from the appearance of bimodality and anisotropic
expansion in time-of-flight images. At this point, the temperature
is 64~nK and total particle number is $5 \times 10^5$. An almost
pure condensate with $10^5$ atoms was obtained after another 2.5~s.
In this evaporation process, the mean truncation parameter is
calculated to be $\bar{\eta}=\langle U/k_B T\rangle = 6.5(3)$, the
evaporation efficiency is
$\bar{\gamma}_{ev}=-\log(\phi/\phi_0)/\log(N/N_0)=3.4$. We observe
an increasing collision rate and accelerating evaporation,
indicating achievement of runaway evaporation; see Fig.~2.

An alternative evaporation path is developed to minimize the time to
reach Cs BEC. After a shorter magnetic field ramping process of 1~s,
we ramp the field gradient from $38.9$~G/cm at $t=0$ to $41.3$~G/cm
in 0.5~s and then to 43.5~G/cm in another 1.5~s. Here we reach BEC
in as short a period as 1.8~s forced evaporation. Another 1~s
evaporation allows us to obtain 4$\times 10^4$ atoms in an almost
pure condensate, see Fig.~2 (d). Despite the rapid increase of phase
space density, the collision rate actually decreases by $25\%$ at
the end of evaporation. The truncation parameter and evaporation
efficiency are $\bar{\eta}=4.6$ and $\bar{\gamma}_{ev}=1.9$,
respectively.

Throughout both evaporation processes, the peak density is moderate,
$n < 1.5\times 10^{13}$cm$^{-3}$. The collision loss rate,
determined from the three-body recombination process
\cite{kraemer2006}, is below $1/40$~s at 20.8~G. Trap loss from
collisions is negligible in the following discussion.

To understand the advantage of the trap-tilting scheme, we analyze
evaporative cooling in a model potential. We combine the magnetic
gradient potential and the gravitational potential as $-\gamma mgz$,
where $\gamma=B'/B'_c-1$. The total potential $V(x,y,z)$ can be
modeled as

\begin{eqnarray} \label{potential}
V=-\frac{U_o}2[e^{-2(x^2+z^2)/w^2}+e^{-2(y^2+z^2)/w^2}]-\gamma mgz,
\end{eqnarray}

\noindent where the first two terms come from the two horizontal
trapping beams, the last term is the tilt potential. Here, we assume
the two beams have the same beam waist $w$ and peak light shift
$U_0/2$ for convenience.

\begin{figure}
\includegraphics[width=3.5in]{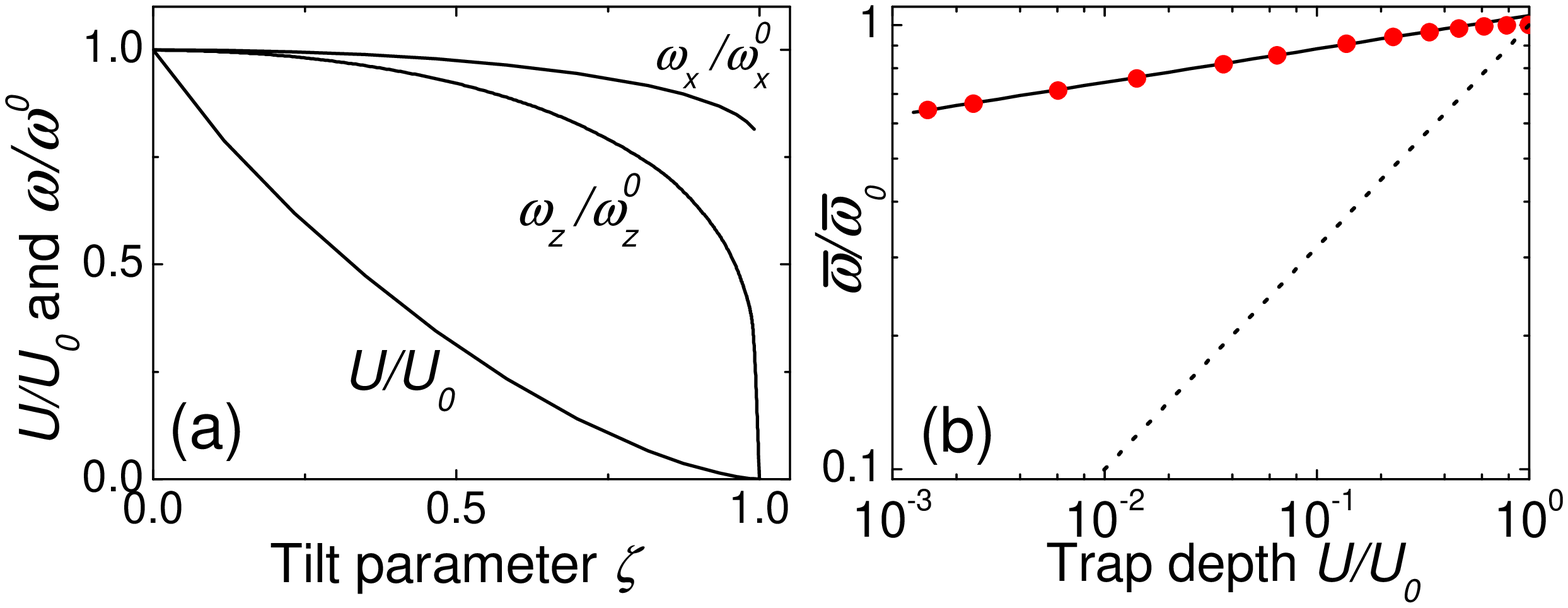}
\caption{Depth and oscillation frequency of a tilted trap. (a) shows
the calculated normalized trap depth and frequencies $\omega_z$ and
$\omega_x=\omega_y$ as a function of the tilt $\zeta$, based on
Eq.~(1). In (b), mean trap frequencies are plotted against the trap
depth for a tilted  trap (solid dots) and for a weakened trap
(dotted line). The solid line shows a polynomial fit to the mean
frequency, see Eq.~(2).} \label{fig:2chan_fig1}
\end{figure}

We introduce the tilt parameter $\zeta=e^{1/2}\gamma mgw/2U_0$ to
parameterize the trap depth $U$ and trap frequencies
$\omega_{x,y,z}$. Using Eq.~(1), the trap depth and frequencies are
evaluated as a function of $\zeta$, as shown in Fig.~3 (a). All
quantities are normalized to those of an untilted potential, where
the trap depth is $U_0$, and the trap frequencies $\omega^0_z=\sqrt
2\omega^0_x=\sqrt 2\omega^0_y=\sqrt{4U_0/mw^2}$. Note that the trap
is unstable when $\zeta\ge1$. In the range of $10^{-3}<U/U_0<1$, the
geometric mean of the trap frequencies
$\bar{\omega}=(\omega_x\omega_y\omega_z)^{1/3}$ varies with the trap
depth approximately as, see Fig.~3 (b),

\begin{equation} \label{nu}
\bar{\omega}/\bar{\omega_0}\approx1.05(U/U_0)^{0.075(1)},
\end{equation}

\noindent where $\bar{\omega_0}$ is the mean frequency of an
untilted trap.


The key to fast, runaway evaporation in a tilted trap lies in the
gentle, almost negligible weakening of the trap confinement when the
trap depth decreases. As the trap depth reduces by a factor of 100,
the trap frequency only decreases by $45\%$ in the $z-$direction and
$14\%$ in the other two directions. This should be contrasted to the
trap-weakening method, which reduces trap frequencies by a factor of
10 under the same condition. In general, a weakening trap with
$\bar{\omega}\propto U^{\nu}$ and $\nu=0.5$ shows a much stronger
dependence on the trap depth than the tilting trap with $\nu=0.075$.

Collision rate in a harmonic trap depends on the particle number,
trap frequencies and temperature. Assuming $\eta=U/k_B T
> 6$ is a constant, we have

\begin{eqnarray} \label{potential}\nonumber
\Gamma & \propto & N \bar{\omega}^3 T^{-1} \\
       & \propto & U^{1/\alpha}U^{3\nu}U^{-1} \equiv U^{\beta},
\end{eqnarray}

\noindent where $\beta=1/\alpha+3\nu-1$,  $\alpha>0$ parameterizes
the temperature decrease by evaporating one atom
\cite{ketterle1996}, namely,

\begin{equation} \label{potential}\nonumber
\alpha=\frac{d\log T}{d\log N}=\frac{\eta+\kappa-3}{3-3\nu},
\end{equation}

\noindent and $\kappa>0$ depends on the dimension of evaporation as
discussed below.



The condition for runaway evaporation is given by $\beta<0$. For the
trap weakening scheme with $\nu=1/2$, $\beta$ is positive for all
$\eta$. Runaway evaporation is thus impossible. For the tilting
scheme with $\nu=0.075$, the exponent $\beta$ is negative when
$\eta+\kappa>6.58$, suggesting runaway evaporation with increasing
collision rate is possible.

\begin{figure}\vspace{10pt}
\includegraphics[width=2.2 in]{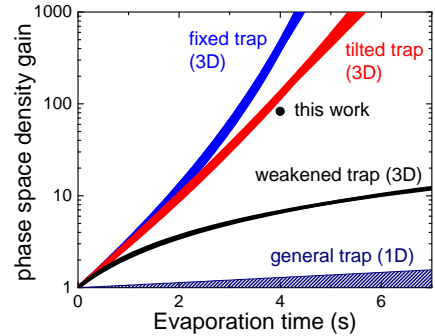}
\caption{Evaporation speed: experiment (4~s path, circle dot) and
models. We assume an initial collision rate of $\Gamma_0$=133~/s,
$\eta=\bar{\eta}=6.2\sim6.8$ and no collision loss. Shaded area
covers 1D evaporation region with $0\leq\nu\leq1$ and all possible
$\eta$.} \label{fig:2chan_fig1}
\end{figure}

Time evolution of the phase space density $\phi(t)$ can be derived
based on standard evaporation theory \cite{ketterle1996}. Assuming
energetic atoms can leave the sample in all directions, we have
$\phi(t)=\phi(0)(1+ \lambda_{3D}\alpha\beta\Gamma_0 t)^{2/\beta-1}$
and $\kappa_{3D}=(\eta-5)/(\eta-4)$ \cite{Luiten1996,ohara2001}.
Here $\Gamma_0$ is the initial collision rate and
$\lambda_{3D}=(\eta-4)e^{-\eta}/\sqrt2$ \cite{Luiten1996,ohara2001}
is the fraction of collisions producing an evaporated atoms. Here we
see that a negative $\beta<0$ leads to a faster-than-exponential
growth of the phase space density, which eventually diverges at time
$t=(-\lambda_{3D}\alpha\beta\Gamma_0)^{-1}$. We compare the models
and our experiment result in Fig.~4. To reach the same final phase
space density, the trap-tilting scheme would require a much shorter
evaporation time than the weakening scheme. For comparison, a
potential with fixed trap frequency ($\nu=0$), e.g., radio-frequency
based evaporation in magnetic traps, permits an even stronger
runaway effect, see Fig.~4.

Remarkably, the performance of our evaporation is consistent with
the 3D evaporation model. The consistency of our evaporation speed
with the 3D model is somewhat surprising. In a strongly tilted trap
where hot atoms can only escape the trap in the tilted direction, it
is generally expected that the evaporation will exhibit performance
consistent with one dimensional evaporation. In 1D evaporation,
energetic atoms are selected only by their velocity in the tilted
direction. Integration over the Boltzmann distribution on only this
degree of freedom then results in a reduction of the evaporation
rate by a factor of $4\eta$ \cite{surkov1996} and thus
$\lambda_{1D}=\lambda_{3D}/4\eta$. Performance of 1D evaporation for
all possible $\eta$ is shown in the shaded area in Fig.~4. Our
experiment result apparently permits evaporation speed much faster
than the 1D prediction.

We suspect 3D-like evaporation in a tilted trap results from the
inseparability of the potential and the existence of a saddle point
located at the rim of the potential barrier. Both features, together
with the similar trap frequencies in all directions, can lead to
stochastic single particle motion \cite{surkov1996}. When atoms with
sufficiently high energy are created by collisions, stochastic
motion allow them to efficiently find escape trajectories. If the
energetic atoms have a high probability to escape, regardless of
their initial direction of motion, evaporation is effectively three
dimensional \cite{surkov1996}. In realistic models, stochastization
may also be induced by the intensity irregularity of the trapping
laser beams.

\begin{figure}\vspace{10pt}
\includegraphics[width=2.2 in]{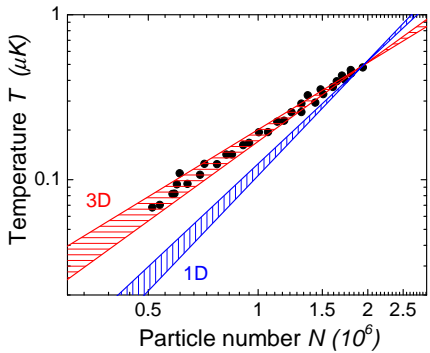}
\caption{(color online) Temperature and particle number dependence.
Based on the 4~s evaporation data (solid circle), the temperature
shows a polynomial dependence on particle number $T\propto
N^{\alpha}$, with $\alpha=1.46(2)$. For comparison, predictions from
3D (red) and 1D (blue) evaporation models are shown using our
experiment initial condition and $\eta=6.5\pm0.3$.}
\label{fig:2chan_fig1}
\end{figure}

To further investigate the ``dimension of evaporation'' in a tilted
trap, we come back to $\eta+\kappa$, which parameterizes the energy
removal by evaporating a single atom, or $\eta+\kappa=-(k_B
T)^{-1}dE/dN$. For 3D evaporation, we expect
$\kappa_{3D}=(\eta-5)/(\eta-4)$, which is $\kappa_{3D}=0.6(1)$ for
our parameter $\bar{\eta}=6.5(3)$; for 1D evaporation, energy
selectivity applies to the axial, but not the transverse motion,
which has a mean energy of 2~$k_BT$ per particle. Hence, we expect a
higher energy removal per particle with
$\kappa_{1D}=\kappa_{3D}+2=2.6(1)$ for our parameter
\cite{pinkse1998}. Experimentally, we can test these predictions by
evaluating $\alpha=d\log T/d\log N$, which has a simple dependence
on $\kappa$ as shown in Eq.~4. We show in Fig.~5 that our 4~s
evaporation data is excellently fit to the polynomial function with
$\bar{\alpha}=1.46(2)$. Using Eq.~4, we derive $\kappa=0.6(3)$,
which is consistent with the 3D value and confirms the 3D nature of
the trap-tilt based evaporation.




This work is supported under ARO Award W911NF0710576 with funds from
the DARPA OLE Program, the NSF-MRSEC program under DMR-0213745 and
the Packard foundation. We thank Jeffrey Gebhardt, Robert Berry and
Selim Jochim for the technical assistance in the early stage of the
experiment.

\end{document}